\documentclass[aps,amsmath,notitlepage,twocolumn,amssymb,prb,letterpaper]{revtex4-1}

\usepackage{graphicx}
\usepackage{dcolumn}
\usepackage{bm}
\usepackage[colorlinks=true,citecolor=red,urlcolor=blue]{hyperref}
\usepackage{wasysym}
\usepackage{stmaryrd}
\usepackage{verbatim}

\begin{document}

\title{Theory of quantum kagome ice and vison zero modes}

\author{Yi-Ping Huang}
\author{Michael Hermele}

\affiliation{Department of Physics, University of Colorado, Boulder, Colorado
80309, USA}
\affiliation{Center for Theory of Quantum Matter, University of Colorado, Boulder, CO 80309, USA}
\date{\today}

\begin{abstract}
	We derive an effective $Z_2$ gauge theory to describe the quantum kagome
	ice (QKI) state that has been observed by Carrasquilla \emph{et. al.} in Monte Carlo studies of the $S = 1/2$ kagome XYZ model
	in a Zeeman field.  The numerical results on QKI are consistent with, but do
	not confirm or rule out, the hypothesis that it is a $Z_2$ spin liquid.  Our effective theory allows us to
	explore this hypothesis and make a striking prediction for future numerical studies, namely that symmetry-protected vison zero modes arise at lattice disclination defects, leading to a Curie defect term in the spin susceptibility, and a characteristic $(N_{dis} - 1) \ln 2$ contribution to the entropy, where $N_{dis}$ is the number of disclinations.  Only the $Z_2$ Ising symmetry is required to protect the vison zero modes.  This is remarkable because a unitary $Z_2$ symmetry cannot be responsible for symmetry-protected degeneracies of local degrees of freedom.  We also discuss other signatures of symmetry fractionalization in the $Z_2$ spin liquid, and phase transitions out of the $Z_2$ spin liquid to nearby ordered phases.
\end{abstract}

\maketitle

\section{Introduction}
\label{sec:intro}

Quantum spin liquids (QSLs) are remarkable zero-temperature phases of insulating
spin systems.\cite{Anderson1973,Anderson1987,Savary2016}  These states lack any sort of symmetry-breaking order, but
instead exhibit long-range quantum entanglement.  Some QSLs are stable phases
with gapless excitations, while others are gapped and topologically ordered,
supporting fractional excitations, as in fractional quantum Hall liquids.  Over
the last several years, a number of candidate materials for gapless QSLs have
emerged (see~[\onlinecite{Savary2016}] and references therein).
Recent Knight shift\cite{Fu2015} and inelastic neutron scattering\cite{Han2016} measurements suggest a gapped spin liquid ground state in
ZnCu$_3$(OD)$_6$Cl$_2$,  but interpretation of these results is complicated by significant impurity effects, while other measurements point to a gapless state.  It remains an important problem to find candidate materials for gapped QSLs.

In a closely related development, numerical studies of simple and
fairly realistic quantum spin models have found evidence for two types of gapped
QSLs, namely $Z_2$ QSLs,\cite{chakraborty89,read91,wen91,balents99,senthil00,moessner01a, moessner01b,Balents2002,kitaev03} and chiral spin liquids.\cite{Kalmeyer1987,Wen1989}  There is evidence for a $Z_2$ QSL
in the $S=1/2$ kagome Heisenberg
antiferromagnet,\cite{Yan2011,Depenbrock2012,Jiang2012,Mei2016} although there are also contrary indications that the
ground state may be gapless.\cite{Iqbal2011,Iqbal2013,Iqbal2014,Liao2016}  In the same model, a chiral spin liquid phase arises upon adding second and third neighbor interactions, with or without XXZ anisotropy.\cite{He2014,Gong2014,Gong2015}  Recently, in the $S=1/2$ $J_1-J_2$ triangular Heisenberg antiferromagnet, density matrix renormalization group studies have found evidence of a gapped spin liquid,\cite{Zhu2015,Hu2015} although a variational wave function approach favors a gapless spin liquid.\cite{Iqbal2016}  These works  raise the
prospects for finding gapped QSLs in real materials, and provide clues where to
look for such states. However, especially given that gapped QSLs are not conclusively established in some of these models, it continues to be important to identify simple, fairly realistic candidate models for gapped QSLs.

\begin{figure}[t]
	\includegraphics[width=0.8\columnwidth]{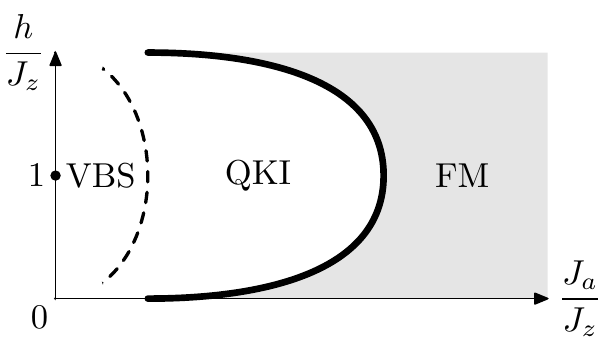}
	\caption{
	Schematic zero-temperature phase diagram of the XYZh model, based on the quantum Monte Carlo results of
		Ref.~\onlinecite{Carrasquilla2015}, showing quantum kagome ice (QKI), ferromagnetic (FM) and valence bond solid (VBS) states.  Only $h > 0$ is shown, as the phase diagram is symmetric under $h \to -h$. At small $J_a / J_z$,
		the system can be mapped to a honeycomb lattice quantum dimer model where VBS order is expected,\cite{Moessner2001} although VBS order was not observed in Ref.~\onlinecite{Carrasquilla2015}, perhaps due to a small temperature scale or problems with equilibration at small $J_a$.  The phase transition from QKI to FM was found to be first-order.}
	\label{fig:phasediagram}
\end{figure}

In an exciting addition to this body of work, Carasquilla, Hao and Melko (CHM) have identified a gapped, quantum disordered phase in a $S = 1/2$ XYZ model on the kagome lattice in a $z$-axis Zeeman magnetic field (XYZh model).\cite{Carrasquilla2015}  CHM proposed this state, dubbed quantum kagome ice, to be a gapped $Z_2$ QSL.

The XYZh model has potential relevance to $f$-electron pyrochlore magnets where effective spin-1/2 degrees of freedom transform not as magnetic dipoles, but instead as dipolar-octupolar Kramers doublets.\cite{Huang2014}  Together with G. Chen, we showed that such systems are described by a XYZ model, which was argued to be particularly relevant for A$_2$B$_2$O$_7$ pyrochlores with A $=
$ Nd;\cite{Huang2014} experiments have found evidence for dipolar-octupolar doublets in some such systems.\cite{Watahiki2011,Hatnean2015,Lhotel2015,Xu2015}  Following prior work on the ``kagome ice'' state of classical spin ice pyrochlores,\cite{Katsuhira2002,Wills2002,Moessner2003,hiroi2003,Sakakibara2003,Isakov2004,tobata2006,Macdonald2011}  CHM noted that the pyrochlore XYZ model descends to the XYZh model on approximately decoupled kagome layers upon applying a magnetic field.

In more detail, CHM considered the Hamiltonian
\begin{eqnarray}
	\mathcal{H}_{\text{XYZh}}&&= \sum_{\langle
	\boldsymbol{r},\boldsymbol{r}'\rangle}
	J_zS^z_{\boldsymbol{r}}S^z_{\boldsymbol{r}'}
	-h\sum_{\boldsymbol{r}}S^{z}_{\boldsymbol{r}}\\
	-&&\sum_{\langle
	\boldsymbol{r},\boldsymbol{r}'\rangle}\left[\frac{J_{\perp}}{2}\left(S^+_{\boldsymbol{r}}S^{-}_{\boldsymbol{r}}+h.c.\right)+\frac{J_{a}}{2}\left(
	S^+_{\boldsymbol{r}}S^{+}_{\boldsymbol{r}}+h.c.\right)\right]\nonumber
\end{eqnarray} where $J_z > 0$, $\boldsymbol{r}$ labels Kagome lattice sites, and $\langle \boldsymbol{r}, \boldsymbol{r}' \rangle$ denotes nearest-neighbor bonds.  CHM set $J_{\perp} = 0$ and used quantum Monte Carlo to obtain the phase diagram as a function of $J_a / J_z$ and $h / J_z$, finding two ``lobes'' of QKI centered at $h / J_z = \pm 1$, as shown in Fig.~\ref{fig:phasediagram}

CHM examined various candidate orders in the QKI state and concluded that it lacks symmetry-breaking order.  Moreover, following prior works,\cite{Nikolic2005,Savary2012,Lee2012}
they showed that $\mathcal{H}_{\text{XYZh}}$ can be exactly rewritten as a ${\rm U}(1)$ gauge theory, with the $J_a$ term a pair-hopping of spinons that can lead to condensation of spinon pairs and thus to a $Z_2$ QSL.  Based on this insight, CHM described how to obtain this state within a gauge mean-field treatment.\cite{Savary2012}

The results of CHM are consistent with the hypothesis that quantum kagome ice is a $Z_2$ QSL, but this has not been directly confirmed or ruled out.  No Lieb-Schultz-Mattis type theorem\cite{Lieb1961,Oshikawa2000,Hastings2004} is believed to hold for the XYZh model, so that a trivial quantum paramagnet is expected to be a possible ground state\footnote{Our effective theory allows us to confirm this expectation.} and is also consistent with the results of CHM.  It is therefore important to devise signatures that can distinguish the $Z_2$ QSL and trivial paramagnet, as well as other possible states.

In this paper, we derive an effective gauge theory of QKI as a $Z_2$ QSL, study its properties, and use it to make a striking prediction that we expect can be tested in future quantum Monte Carlo studies.  In particular, we show that lattice disclination defects host vison zero modes, \emph{i.e.} there is no energy cost to insert a vison at a disclination.  The resulting degeneracies only require the $Z_2$ Ising symmetry of the XYZh model for their protection, which is remarkable because a unitary $Z_2$ symmetry cannot protect degeneracies of local degrees of freedom.  The vison zero modes lead to a Curie spin susceptibility localized at the defects.  In addition, in a system without boundary where $N_{dis}$ disclinations host vison zero modes, there are $2^{N_{dis}} / 2$ degenerate states associated with the zero modes, where the factor of $1/2$ comes from the global constraint of an even number of visons.  The resulting $(N_{dis} - 1) \ln 2$ contribution to the entropy directly distinguishes vison zero modes from local doublets bound to disclinations, which would have a degeneracy of $2^{N_{dis}}$.  We also discuss other possible signatures related to the symmetry properties of spinon and vison excitations, both within the $Z_2$ spin liquid phase and at phase transitions to nearby symmetry-breaking phases.

In Sec.~\ref{sec:derivation} we derive the effective $Z_2$ gauge theory, starting from an exact rewriting of the XYZh model as a ${\rm U}(1)$ gauge theory.\cite{Carrasquilla2015,Nikolic2005,Savary2012,Lee2012}  We then discuss the role of symmetry in the $Z_2$ QSL (Sec.~\ref{sec:symmetry}).  We find that the spinon has non-trivial symmetry fractionalization, while the symmetry fractionalization of the vison is trivial; the computation of the symmetry fractionalization is discussed in Appendix~\ref{app:frac}.  Section~\ref{sec:disclinations} describes the vison zero modes at lattice disclinations their signatures in spin susceptibility and entropy.  Other properties of the $Z_2$ QSL, including phase transitions to nearby phases, are discussed in Sec.~\ref{sec:other}, and the paper concludes with a brief discussion (Sec.~\ref{sec:discussion}).

We would like to note other current work on the theory the spin liquid state in the XYZh model, using an approach complementary to our own.\cite{Chen2016}

\begin{figure}[t]
	\includegraphics[width=\columnwidth]{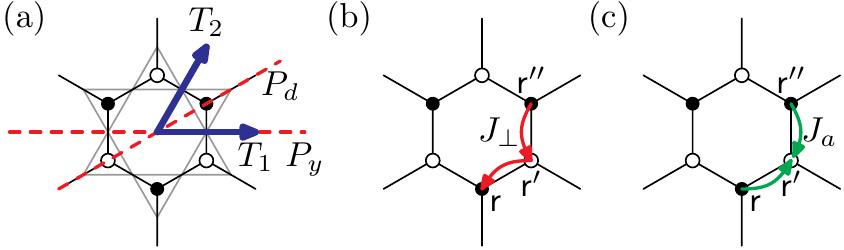}
	\caption{(a) The sites of the kagome lattice, where spins of the XYZh model reside, are identified with nearest-neighbor links of the honeycomb lattice.  Honeycomb sites, which correspond to kagome triangles, naturally divide into A and B sublattices, shown as open and closed circles, due to the bipartite nature of the honeycomb lattice.  The generators of the $p6m$ space group are shown, with $P_d$ and $P_y$ reflections (dashed lines) and $T_1$ and $T_2$ translations (thick arrows).  (b) and (c) illustrate the hopping processes in the ${\rm U}(1)$ gauge theory that correspond to $J_{\perp}$ and $J_a$ terms, respectively.}
	\label{fig:geometry}
\end{figure}

\section{Derivation of effective gauge theory}
\label{sec:derivation}

Our effective gauge theory is based on an exact rewriting of the XYZh model as a ${\rm U}(1)$ gauge theory.  Before getting into details, we motivate the rewriting by considering the classical limit $J_{\perp} = J_{a} = 0$ and $h = J_z$, where the ground states are configurations of $S^z_{\boldsymbol{r}}$ with two spins up and one spin down on every triangle.  Kagome sites correspond to nearest-neighbor links of the dual honeycomb lattice, while kagome triangles correspond to honeycomb sites (Fig.~\ref{fig:geometry}a).  We can view  up-up-down spin configurations as dimer coverings of the honeycomb lattice, associating down spin (up spin) with presence (absence) of a dimer.  Moving slightly away from the classical case by allowing $0 \neq J_a, J_{\perp} \ll J_z$, we obtain a honeycomb quantum dimer model, which is a ${\rm U}(1)$ gauge theory.  

Now we proceed to rewrite the XYZh model as a ${\rm U}(1)$ gauge theory, without making any assumptions about the size of the various couplings in the Hamiltonian.  This rewriting follows CHM,\cite{Carrasquilla2015} who in turn followed Refs.~\onlinecite{Nikolic2005,Savary2012,Lee2012}.  We first introduce the Hilbert space and operators of the gauge theory, and then describe their relationship to the Hilbert space and spin operators of the XYZh model.  We label the sites of the honeycomb lattice by sans serif letters ${\sf r}$.  On each honeycomb link we place a ${\rm U}(1)$ quantum rotor, with number $e_{\sf r r'}$ that will be the electric field, and phase $a_{\sf r r'}$ that will be the vector portential.  On the same link, these operators satisfy the commutation relation $[a_{\sf r r'}, e_{\sf r r'}] = i$, and we define $e_{\sf r' r} = - e_{\sf r r'}$ (similarly for $a_{\sf r r'}$).  On honeycomb sites we also place ${\rm U}(1)$ quantum rotors with number $n_{\sf r}$ and phase $\theta_{\sf r}$, satisfying $[\theta_{\sf r}, n_{\sf r}] = i$.  The site degrees of freedom are matter fields carrying the ${\rm U}(1)$ gauge charge.  To fully specify the gauge theory Hilbert space we need to specify the Gauss' law constraint, which we take to be
\begin{equation}
(\operatorname{div} e)_{\sf r} = 2 \eta_{\sf r} + Q_{\sf r} \text{,} \label{eqn:gauss}
\end{equation}
where $Q_{\sf r} \equiv n_{\sf r}$ is the gauge charge at ${\sf r}$, $2 \eta_{\sf r}$ is a static background charge, and we have defined $\eta_{\sf r}$ to be $1$ ($-1$) for ${\sf r}$ in the A (B) sublattice.  The lattice divergence is defined by $(\operatorname{div} e)_{\sf r} = \sum_{{\sf r}' \sim {\sf r}} e_{\sf r r'}$, where the sum is over the three neighbors of ${\sf r}$.

The gauge theory Hilbert space is identical to that of the spin model, if we impose the additional ``hardcore'' constraint $e_{\sf r r'} = 0,1$, with $\sf r$ in the A sublattice.  Then we impose the relation
\begin{equation}
e_{\sf r r'} = \eta_{\sf r} ( S^z_{\sf r r'} + 1/2 ) \text{,}
\end{equation}
where we take $S^z_{\sf r r'} \equiv S^z_{\sf r' r}$.  This says that Ising spin configurations are the same as electric field configurations.  Gauss' law then determines $Q_{\sf r}$, giving
\begin{equation}
Q_{\sf r} = \eta_{\sf r} \big( \sum_{\boldsymbol{r} \in \triangle} S^z_{\boldsymbol{r}} - 1/2 \big) \text{,}
\end{equation}
where $\triangle$ is the triangle whose center is $\sf r$.  We see that $Q_{\sf r}$ is zero for triangles in an up-up-down spin configuration, and measures the deviation of the total spin on a triangle from $1/2$. In fact, we included the background charge $2 \eta_{\sf r}$ in Gauss' law in order to make this property hold.

\begin{figure}[t]
	\includegraphics[width=0.5\columnwidth]{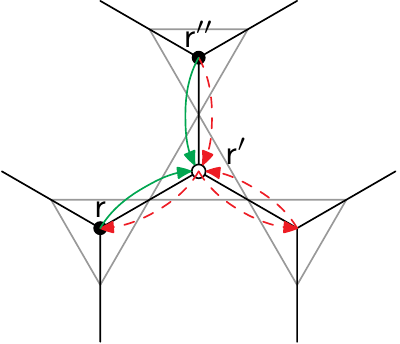}
	\caption{Illustration of how $J_a$ and $J_{\perp}$ terms can combine to give nearest-neighbor charge-two hopping.  Three different $J_a$ coordinated hopping processes are shown.  Two of these are shown with red dashed arrows, one with green solid arrows.  Acting in succession with the two red/dashed processes gives a charge-one hopping from ${\sf r}''$ to ${\sf r}$, the same process as the $J_{\perp}$ term.   Combining this with the green/solid process then gives a charge-two hopping from ${\sf r}''$ to ${\sf r}'$.}
	\label{fig:generating-charge-two-hopping}
\end{figure}

To complete the mapping between the gauge theory and spin model, we write
\begin{equation}
S^{+}_{\sf r r'} = \exp\big( i \eta_{\sf r} (\theta_{\sf r} - \theta_{\sf r'} + a_{\sf r r'} ) \big) \text{,}
\end{equation}
where again we take $S^{+}_{\sf r r'} \equiv S^{+}_{\sf r' r}$.  This formula has a simple interpretation, namely that $S^+_{\sf r r'}$ hops a gauge charge between two neighboring sites of the honeycomb lattice.

Taking $h = J_z$ for simplicity, which puts us at the center of one of the lobes of QKI, in terms of the gauge theory degrees of freedom the XYZh Hamiltonian becomes
\begin{eqnarray}
\mathcal{H}_{\text{gauge}} &=& \frac{J_z}{2} \sum_{\sf r} n_{\sf r}^2 + U \sum_{{\sf r} \in A} \sum_{{\sf r}' \sim {\sf r}} ( e_{\sf r r'} - 1/2)^2 \label{eqn:hgauge} \\
&-& J_{\perp} \sum_{\langle\langle {\sf r , r''} \rangle \rangle} \cos(\theta_{\sf r} - \theta_{\sf r''} + a_{\sf r r'} + a_{\sf r' r''} ) \nonumber \\
&-& J_a  \sum_{\langle\langle {\sf r , r''} \rangle \rangle} \cos(2 \theta_{\sf r'} - \theta_{\sf r} - \theta_{\sf r''} + a_{\sf r' r} + a_{\sf r' r''} ) \text{.}  \nonumber
\end{eqnarray}
The sum in the latter two terms is over pairs of next-nearest neighbor honeycomb sites ${\sf r, r''}$, with ${\sf r'}$ the site ``in between'' ${\sf r}$ and ${\sf r''}$ as shown in Fig.~\ref{fig:geometry}b,c.  In order to obtain a useful effective theory, we have softened the hardcore constraint on electric fields with the $U$ term, which restores this constraint in the limit $U \to \infty$, where the original spin model is recovered.  We see that the $J_{\perp}$ term is a next-nearest neighbor hopping of gauge charges (Fig~\ref{fig:geometry}b).  The $J_a$ term is a coordinated hopping, which moves unit charges from sites ${\sf r}$ and ${\sf r}''$ together into site ${\sf r'}$ (Fig.~\ref{fig:geometry}c).

The coordinated $J_a$ hopping can loosely be thought of as motion of a charge-two object.  As was suggested for very similar ${\rm U}(1)$ gauge theories on the pyrochlore lattice,\cite{Lee2012} and in the present context by CHM,\cite{Carrasquilla2015} it is thus reasonable that $J_a$ may drive condensation of a charge-two field, while leaving single charge excitations gapped.  Such a condensation breaks the ${\rm U}(1)$ gauge structure down to $Z_2$,\cite{Fradkin1979} thus leading to a $Z_2$ spin liquid.

We note that a nearest-neighbor charge-two hopping can indeed be generated from the $J_a$ hopping process, or from $J_a$ and $J_{\perp}$ processes together, as illustrated in Fig.~\ref{fig:generating-charge-two-hopping}.  This motivates us to introduce a charge-two field with number $N_{\sf r}$ and phase $\Theta_{\sf r}$, which represents a bound state of two unit $\theta_{\sf r}$ gauge charges.  We add the following terms to the Hamiltonian:
\begin{eqnarray}
\delta {\cal H} &=&  u_2 \sum_{\sf r} N_{\sf r}^2 - \Delta \sum_{\sf r} \cos(\Theta_{\sf r} - 2 \theta_{\sf r} )  \\
&-& t_2 \sum_{\langle {\sf r r'} \rangle} \cos(\Theta_{\sf r} - \Theta_{\sf r'} + 2 a_{\sf r r'} )
- K \sum_{\hexagon} \cos( \nabla \times  a ) \text{.} \nonumber
\end{eqnarray}
The first term is a repulsive interaction for the new charge-two field.
The second term corresponds to a process where two unit charges convert to a single double charge, and the third term is nearest-neighbor hopping of double charges.  The last term is a Maxwell term for the ${\rm U}(1)$ gauge field, where the sum is over honeycomb hexagons and $\nabla \times a$ is the discrete line integral of $a_{\sf r r'}$ around the perimeter of a hexagon.  The Maxwell term suppresses ${\rm U}(1)$ gauge fluctuations, and is the leading dynamical term generated in degenerate perturbation theory when $J_{\perp} , J_a \ll J_z$.  While we do not work in that limit, the fact that the Maxwell term is generated there makes it reasonable to add it explicitly to our effective Hamiltonian.  For consistency, we also redefine $Q_{\sf r} \equiv n_{\sf r} + 2 N_{\sf r}$ in Eq.~(\ref{eqn:gauss}).

We now take the $\Delta$ and $t_2$ terms in $\delta {\cal H}$ to be large.  The $t_2$ term drives condensation of the charge-two field, while $\Delta$ is taken large for convenience.  Provided $K$ is sufficiently large, this drives the system into the $Z_2$ QSL phase and allows us to obtain an effective gauge theory describing it.

Taking $t_2$ large and treating the cosine as a constraint, we have
\begin{equation}
a_{\sf r r'} = \frac{1}{2} \Theta_{\sf r'} - \frac{1}{2} \Theta_{\sf r} + \alpha_{\sf r r'} \text{,}  \label{eqn:t2-constraint}
\end{equation}
where $\alpha_{\sf r r'}$ takes values $0, \pi$.  There is an ambiguity in multiplying a ${\rm U}(1)$ phase by $1/2$, which is the same as the ambiguity in defining the square root for complex numbers.  We pick a branch by associating a ${\rm U}(1)$ phase $\phi$ with the corresponding real number lying in the interval $[-\pi, \pi)$, for which multiplication by $1/2$ is defined in the usual way.

The other effect of treating the $t_2$ term as a constraint is that only operators commuting with the term survive in the low-energy Hilbert space.  In particular, $e_{\sf r r'}$ does not commute with the constraint, but
\begin{equation}
\sigma^x_{\sf r r'} \equiv \exp(i \pi e_{\sf r r'} ) 
\end{equation}
does, and becomes the $Z_2$ electric field.  We also define the $Z_2$ vector potential $\sigma^z_{\sf r r'} \equiv \exp(i \alpha_{\sf r r'})$, which anticommutes with $\sigma^x_{\sf r r'}$ on the same link, justifying the Pauli matrix notation.  

Similarly, taking $\Delta$ large gives the constraint
\begin{equation}
\theta_{\sf r} = \frac{1}{2} \Theta_{\sf r} + t_{\sf r} \text{,} \label{eqn:delta-constraint}
\end{equation}
where $t_{\sf r} = 0, \pi$ and we define $\tau^z_{\sf r} \equiv \exp(i t_{\sf r})$.  We also introduce $\tau^x_{\sf r} \equiv \exp(i \pi n_{\sf r})$, which, unlike $n_{\sf r}$ or $N_{\sf r}$, commutes with the $\Delta$ term.

To write the low-energy effective Hamiltonian, those terms commuting with the constraints can straightforwardly be simplified using Eqs.~(\ref{eqn:t2-constraint},\ref{eqn:delta-constraint}).  Terms not commuting with the constraint need to be replaced by new terms acting within the low-energy Hilbert space.  Rather than try to determine those terms systematically, we simply write down the simplest such terms consistent with symmetry (taking input from Sec.~\ref{sec:symmetry}), and use physical arguments to further constrain the corresponding parameters.  The effective Hamiltonian is
\begin{eqnarray}
{\cal H}_{\text{eff}} &=& - K \sum_{p} B_p
- J \sum_{\langle\langle {\sf r , r''} \rangle \rangle} \tau^z_{\sf r} \sigma^z_{\sf r r'} \sigma^z_{\sf r' r''} \tau^z_{\sf r''}  \nonumber \\
&-& v \sum_{\langle {\sf r r'} \rangle} \sigma^x_{\sf r r'} - u \sum_{\sf r} \tau^x_{\sf r} \text{,} \label{eqn:heff}
\end{eqnarray}
where the first sum is over hexagonal plaquettes $p$ and $B_p \equiv \prod_{{\sf r r'} \in p} \sigma^z_{\sf r r'}$.
The first term is obtained directly from the $K$ term in $\delta {\cal H}$, and the $J$ term from the $J_{\perp}$ and $J_a$ terms of the original Hamiltonian.  The latter two terms are the simplest symmetry-allowed terms giving dynamics to $\sigma^z_{\sf r r'}$ and $\tau^z_{\sf r}$, in accord with the discussion above.  The $Z_2$ gauge constraint is obtained by exponentiating Eq.~(\ref{eqn:gauss}) and is
$\prod_{{\sf r}' \sim {\sf r}} \sigma^x_{\sf r r'} = \tau^x_{\sf r}$.  It should be noted that the background ${\rm U}(1)$ gauge charge $2 \eta_{\sf r}$ has dropped out.

We are free to choose $u, v > 0$ by making unitary transformations $\tau^x \to - \tau^x$  ($\sigma^x \to - \sigma^x$) to change the sign of $u$ ($v$).  Each of these transformations introduces a minus sign into the gauge constraint, which becomes $\prod_{{\sf r}' \sim {\sf r}} \sigma^x_{\sf r r'} = \pm \tau^x_{\sf r}$, with an undetermined sign that we now fix below by a physical argument.

First, we need to describe the excitations of the $Z_2$ spin liquid phase that the model enters when $K$ is sufficiently large compared to the other terms in ${\cal H}_{\text{eff}}$.  This puts the $Z_2$ gauge field in its deconfined phase.  There are two types of gapped excitations:  spinons carrying the $Z_2$ gauge charge, and visons carrying the $Z_2$ gauge flux.  $\tau^x = -1$ ($1$) indicates the presence (absence) of a spinon, so that $u$ controls the spinon gap.  Visons reside on hexagons with $B_p = -1$.

To fix the sign of the gauge constraint, we recall that a VBS state is expected to occur adjacent to the $Z_2$ spin liquid for $J_a, J_\perp \ll J_z$, based on the mapping to the honeycomb quantum dimer model (see Fig.~\ref{fig:phasediagram} and Ref.~\onlinecite{Moessner2001}).  We suppose that this VBS can be accessed by condensation of either the spinons or visons of the $Z_2$ spin liquid.  In the same limit where VBS occurs, spinons correspond to defect triangles that violate the up-up-down constraint, and thus have a large energy gap.  Therefore vison condensation is the only option to access the VBS.

We can integrate out spinons to obtain a pure $Z_2$ gauge theory, keeping only the $K$ and $h$ terms of ${\cal H}_{{\rm eff}}$, with gauge constraint $\prod_{{\sf r}' \sim {\sf r}} \sigma^x_{\sf r r'} = \pm 1$, corresponding to presence ($-1$) or absence ($+1$) of a background gauge charge.  It should be noted that this background charge has no direct connection to the background charge $2 \eta_{\sf r}$ in the ${\rm U}(1)$ gauge theory.  Visons reside on sites of the dual triangular lattice (honeycomb hexagons), and feel a background charge as a $\pi$ flux.  With zero flux, the minimum of the vison dispersion lies at the $\Gamma$ point of the Brillouin zone, and we expect visons to condense at zero momentum if $v$ is made sufficiently large.  This leads to a confined phase without breaking lattice symmetry.  On the other hand, visons hopping in background $\pi$ flux have degenerate dispersion minima at the zone corners ($K$ points), so that lattice symmetries are necessarily broken when large enough $v$ drives their condensation, and the confined phase is a VBS.  Therefore we take the gauge constraint to be
\begin{equation}
\prod_{{\sf r}' \sim {\sf r}} \sigma^x_{\sf r r'} = - \tau^x_{\sf r} \text{.}  \label{eqn:z2gauss}
\end{equation}

It should be noted that the presence of background $Z_2$ gauge charge is a non-universal feature of our effective theory, that in principle can be changed by tuning parameters (although it not clear which parameter to tune in the XYZh model to achieve this).  If $v$ is reduced and eventually made negative, we can make a unitary transformation $\sigma^x_{\sf r r'} \to - \sigma^x_{\sf r r'}$ to again make the coefficient of $\sigma^x$ negative in ${\cal H}_{{\rm eff}}$, and remove the background charge from Gauss' law.  This can be done while remaining within the $Z_2$ spin liquid phase, and can be thought of as simply reversing the sign of the vison hopping matrix element.  From this new point in parameter space of the $Z_2$ spin liquid, it is clearly possible to condense visons at zero momentum and enter a trivial phase.  This shows that a trivial quantum paramagnet is indeed possible in the XYZh model, although to access this phase it may be necessary to add additional symmetry-allowed terms to the Hamiltonian.  We remark that this situation is distinct from that occurring in effective theories for other $Z_2$ spin liquids.  For instance, a gapped $Z_2$ spin liquid in the $S=1/2$ kagome Heisenberg model [with ${\rm SU}(2)$ symmetry] necessarily has a background $Z_2$ gauge charge, which is tied to the odd number of $S=1/2$ moments in each unit cell and to the impossibility of a trivial quantum paramagnet in such a model.\cite{Paramekanti2004}

\section{Symmetry in the $Z_2$ spin liquid}
\label{sec:symmetry}

The symmetry group of the XYZh model is $G= Z_2^I \times Z_2^T \times
p6m$, where $Z_2^I$ (generated by ${\cal I}$) is the Ising spin symmetry given by a $\pi$ rotation about the $z$-axis in spin space, and $p6m$ is the space group of the kagome lattice.  While the Zeeman field $h$ breaks the usual time reversal symmetry for spin systems, the XYZh Hamiltonian does enjoy a modified time reversal symmetry ($Z_2^T$, generated by ${\cal T}$) that leaves both $S^z_{\boldsymbol{r}}$ and $S^{+}_{\boldsymbol{r}}$ invariant; this is the natural time reversal operation if we view the XYZh model as a hardcore boson system.

\begin{table}[htp]
	\centering
	\begin{tabular}{|c|cc|cccc|}
	\hline
	 &$S^z_{\boldsymbol{r}}$ &$S^{\pm}_{\boldsymbol{r}}$ &$\theta_{\sf r}$
	&$n_{\sf r}$ &$a_{\sf rr'}$ &$e_{\sf rr'}$\\
	\hline
	$g$ &$S^z_{g(\boldsymbol{r})}$ &$S^{\pm}_{g(\boldsymbol{r})}$ &$\epsilon_g \theta_{g({\sf r})}$
	&$\epsilon_g n_{g({\sf r})}$ &$\epsilon_g a_{g({\sf r}),g({\sf r'})}$
	&$\epsilon_g e_{g({\sf r}),g({\sf r'})}$\\
	\hline
	$\mathcal{I}$ &$S^z_{\boldsymbol{r}}$ &$-S^{\pm}_{\boldsymbol{r}}$
	&$\theta_{\sf r}$ &$n_{\sf r}$ &$a_{\sf rr'}+\pi$ &$e_{\sf rr'}$\\
	\hline
	$\mathcal{T}$ &$S^z_{\boldsymbol{r}}$ &$S^{\pm}_{\boldsymbol{r}}$ &$-\theta_{\sf r}$
	&$n_{\sf r}$ &$-a_{\sf rr'}$ &$e_{\sf rr'}$\\
	\hline\hline
	& - & - & $\tau^z_{\sf r}$ & $\tau^x_{\sf r}$ & $\sigma^z_{\sf r r'}$ & $\sigma^x_{\sf r r'}$ \\
	\hline
	$g$ & - & - & $\tau^z_{g({\sf r})}$ & $\tau^x_{g({\sf r})}$ & $\sigma^z_{g({\sf r}), g({\sf r'})}$ &  $\sigma^x_{g({\sf r}), g({\sf r'})}$\\
	\hline
	$\mathcal{I}$ & - & - & $\tau^z_{{\sf r}}$ & $\tau^x_{{\sf r}}$ & $-\sigma^z_{{\sf r r'}}$ &  $\sigma^x_{{\sf r r'}}$\\
	\hline
	$\mathcal{T}$ & - & - & $\tau^z_{{\sf r}}$ & $\tau^x_{{\sf r}}$ & $\sigma^z_{{\sf r r'}}$ &  $\sigma^x_{{\sf r r'}}$\\
	\hline
	\end{tabular}
	\caption{Action of symmetry operations $g, {\cal I}, {\cal T}$ on the variables of the spin model and ${\rm U}(1)$ gauge theory (above double line), and $Z_2$ effective gauge theory (below double line).   Here $g$ is an element of the $p6m$ space group
	and $\epsilon_g= +1$ ($-1$) when $g$ preserves (exchanges) the A and B honeycomb sublattices.  The transformations of $\Theta_{\sf r}$ and $N_{\sf r}$ are the same as those of $\theta_{\sf r}$ and $n_{\sf r}$.}
	\label{tab:1}
\end{table}

Table~\ref{tab:1} shows how the variables of the spin model and ${\rm U}(1)$ and $Z_2$ gauge theories transform under symmetry. Because $\theta_{\sf r}$ and $a_{\sf r r'}$ are not gauge invariant, there is a gauge arbitrariness in choosing their symmetry transformations.  We have made particular choices to simplify the discussion of the effective $Z_2$ gauge theory; it is possible to make other gauge-equivalent choices, but this has no effect on the physics and does not lead to different possible effective theories.  The transformations in Table~\ref{tab:1} can be obtained from the definitions of the operations quoted for spin operators, by using the expressions that relate the ${\rm U}(1)$ and $Z_2$ gauge theory variables to spin operators and to one another.

With the symmetry transformations in hand, we can compute the action of symmetry on the spinon and vison excitations of the $Z_2$ spin liquid.  Because these are fractional excitations, their behavior under symmetry is an instance of \emph{symmetry fractionalization}.\cite{Wen2002,Essin2013,Chen2015,XChen2016}  By computing the symmetry fractionalization of the spinons and visons, we characterize the $Z_2$ spin liquid as a symmetry enriched topological (SET) phase,\cite{Wen2002,Essin2013,Mesaros2013,Hung2013} which is a starting point for determining its universal properties tied to symmetry.

To characterize the spinon and vison symmetry fractionalization, we first specify the symmetry group in terms of generators and relations.  We choose generators ${\cal I}$, ${\cal T}$, $P_d$, $P_y$, $T_1$ and $T_2$, where the $p6m$ generators are described graphically in Fig.~\ref{fig:geometry}a.  The generators obey the  relations
\begin{eqnarray}
	\mathcal{I}^2 &=& \mathcal{T}^2= {\cal I} {\cal T} {\cal I} {\cal T} = 1\\
	\left( P_d \right)^2&=& \left( P_y\right)^2=\left( P_dP_y
	\right)^6\nonumber\\
	&=& T_1T_2T_1^{-1}T_2^{-1}=T_1P_yT_1^{-1}P_y= 1\\
	T_2&=& P_dT_1P_d;T_2P_y=P_yT_1T_2^{-1}  \text{.}
\end{eqnarray}
In addition, there are six more relations dictating that ${\cal T}$ and ${\cal I}$ commute with $P_d$, $P_y$ and $T_1$ (it then follows from the other relations that the internal symmetries also commute with $T_2$).  Taken together, these relations completely specify the group multiplication.

We introduce operators ${\cal I}^e$ and ${\cal I}^m$, and similarly for the other generators, giving the action of symmetry on spinons ($e$) and visons ($m$).  These operators obey the same relations up to minus signs, and the pattern of minus signs for all the relations specifies the symmetry fractionalization of the corresponding excitation.  The spinon and vison symmetry fractionalizations are computed in Appendix~\ref{app:frac}.  For the visons, we find that all the relations hold with positive signs; that is, the vison has trivial symmetry fractionalization.  On the other hand, the spinon has non-trivial symmetry fractionalization;  we find
\begin{equation}
{\cal I}^e P^e_y =  - P^e_y {\cal I}^e \text{,} \label{eqn:spinon-nontrivial}
\end{equation}
while all other relations hold with a positive sign.  This means that, acting on spinons, the Ising symmetry anticommutes with space group operations that exchange the A and B sublattices, but commutes with operations not exchanging the sublattices.

There have been many studies of $Z_2$ spin liquids on the kagome lattice with continuous spin symmetry, either ${\rm U}(1)$ spin rotations about the $z$-axis, or full ${\rm SU}(2)$ symmetry.  It is interesting to ask whether the QKI $Z_2$ spin liquid is related to any of these states.  In fact, it is impossible to start with such a state, and obtain the QKI $Z_2$ spin liquid by weak explicit breaking of the continuous spin symmetry down to $Z_2^I$.  This is so because with continuous spin symmetry, the ${\cal I}$ operation can be continuously deformed to the identity, so that ${\cal I}^e$ must commute with all the discrete symmetry generators, which is not consistent with Eq.~(\ref{eqn:spinon-nontrivial}).

\section{Vison zero modes at disclinations}
\label{sec:disclinations}

Here, we consider disclination defects of the crystal lattice, and show that the  $Z_2$ QSL has symmetry-protected zero modes bound to these defects.  These zero modes are visons that cost exactly zero energy as long as Ising symmetry is preserved.  We describe observable signatures of the vison zero modes that can be probed in future quantum Monte Carlo studies.  We note that very similar anyon zero modes at symmetry-flux defects of on-site symmetries, and also at lattice dislocations, have been described previously in Ref.~\onlinecite{Cheng2015}.

\begin{figure}[t]
	\includegraphics[width=0.5\columnwidth]{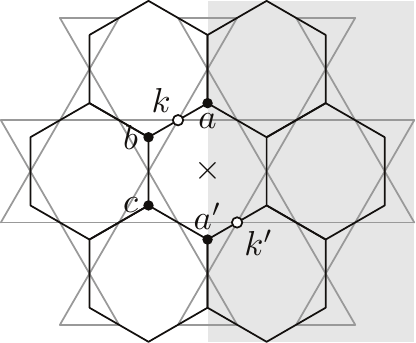}
	\caption{$\pi$ disclination at a hexagon center of the kagome lattice, with the dual honeycomb lattice also shown.  The disclination is a defect where the shaded region is cut out, and sites of the remaining lattice that are related by a $\pi$ rotation about the disclination center are identified.  Equivalently, rather than cut out the shaded region, we can simply identify all sites related by a $\pi$ rotation, such as the kagome sites $k$ and $k'$.  Similarly, the honeycomb site $a$ is identified with $a'$.  The  hexagonal honeycomb plaquette at the disclination center becomes a triangular plaquette with sides $ab$, $bc$, $ca' \simeq ca$.}
	\label{fig:disclination}
\end{figure}

Figure~\ref{fig:disclination} shows a $\pi$ disclination centered at a hexagon.  The disclination is a defect of the lattice where points related by a $\pi$ rotation at the disclination center are identified.  Apart from identifying points related by the $\pi$ rotation, we focus on a special type of disclination where the Hamiltonian density away from the disclination center is left unchanged, \emph{i.e.} the Hamiltonian of the XYZh model on every site and link is the same as in the defect-free system.  Such a disclination preserves the Ising symmetry.    All of our results continue to hold for more generic $\pi$ disclinations, as long as Ising symmetry is preserved, and as long as the Hamiltonian density is unchanged in the far field of the defect.  Our results also hold for any disclination that identifies sites in the A sublattice with sites in the B sublattice (\emph{i.e.}, for $\pm \pi/3$ disclinations), but not for $\pm 2 \pi /3$ disclinations that preserve the bipartite structure of the honeycomb lattice.

We first consider the effect of a single disclination in an infinite plane, using the effective $Z_2$ gauge theory of the $Z_2$ QSL.  We go to the exactly solvable point of ${\cal H}_{{\rm eff}}$ deep within the spin liquid phase, by setting $J = v = 0$.\footnote{At this point, the gauge theory is equivalent to the exactly solvable point of the toric code model on the honeycomb lattice.  See Ref.~\onlinecite{kitaev03} and Appendix~\ref{app:frac}.}  At this point, the exact eigenstates are labeled by eigenvalues of the commuting operators $\tau^x_{\sf r}$ and $B_p$, and the spinon and vison excitations do not propagate.  We observe that all hexagonal plaquettes remain locally unchanged, except for the hexagon $p_{dis}$ at the disclination center, which becomes a closed loop of \emph{three} links.  This implies that $B_{p_{dis}}$ is odd under the Ising symmetry, and in order to preserve Ising symmetry we must set the coupling $K_{dis}$ of this term to zero.  Therefore, putting a vison at the core of the disclination costs zero energy, and we have a pair of degenerate vison / no-vison states.

Remarkably, this vison zero mode is protected by the $Z_2^I$ Ising symmetry; this is unusual because normally a unitary $Z_2$ symmetry cannot lead to symmetry-protected degeneracies.  To see the symmetry protection, consider an effective $2 \times 2$ matrix Hamiltonian for the doublet of vison / no-vison states, $H_{{\rm doublet}} = a_x \sigma^x + a_y \sigma^y + a_z \sigma^z$.  $a_z$ corresponds to $K_{dis}$ and is forbidden by Ising symmetry.  The off diagonal terms $a_x$ and $a_y$ are also forbidden, because no  local operator we might use to perturb the Hamiltonian can create or destroy a vison and flip between $\sigma^z$ eigenstates.  The doublet therefore remains degenerate as long as Ising symmetry is preserved -- no other symmetries are needed for its protection.  While lattice rotation of the defect-free system plays an important role, allowing us to introduce the disclination in the first place, it and other point group symmetries are not needed to protect the zero mode.

We now turn to observable signatures of the vison zero modes.  The degeneracy will be lifted if the Ising symmetry is broken explicitly, because $K_{dis} \neq 0$ is then allowed.  Returning to the XYZh model, this can be achieved by adding a local transverse field
\begin{equation}
H_{{\rm transverse}} = - \sum_{\boldsymbol{r}} h_x(\boldsymbol{r}) S^x_{\boldsymbol{r}} \text{,}
\end{equation}
where $h_x(\boldsymbol{r})$ non-zero only near the disclination center.  This implies that disclinations contribute a Curie term in the temperature dependence of the transverse spin susceptibility $\chi_{xx}(T)$.   Since only the spins near the defect contribute to the Curie  susceptibility, to detect this effect it is sufficient to look at the local susceptibility of spins in some region near the disclination.  Indeed, looking at the local susceptibility is preferable to better separate bulk and impurity contributions to $\chi_{xx}(T)$; away from the disclination, $\chi_{xx}(T)$ goes to a constant as $T \to 0$.  The Curie behavior should be observable within a temperature range $T_{{\rm low}} < T < T_{{\rm gap}}$, where $T_{{\rm gap}}$ is the lowest bulk energy gap of the XYZh model in temperature units, and $T_{{\rm low}}$ corresponds to the energy scale for interactions between vison zero modes on nearby disclinations.  Such interactions require visons to tunnel through the bulk where they are gapped, and thus go to zero exponentially in the separation between disclinations.

The vison zero modes also have an interesting manifestation in entropy as measured by heat capacity.  We consider a finite system without boundary, which has an even number $N_{dis}$ of disclinations.  Na\"{\i}vely we might guess the total degeneracy is $2^{N_{dis}}$, but this is not correct due to the constraint that the total number of visons in the system must be even.  This means that the total degeneracy is in fact $2^{N_{dis} - 1}$.  In principle, this should be observable in quantum Monte Carlo by measuring the heat capacity in a small transverse field, and integrating the resulting Schottky peak to obtain the entropy of $(N_{dis} - 1) \ln 2$ associated with the gapless defect modes.

This latter signature is important, as it differentiates vison zero modes from a collection of local doublets bound to disclinations (\emph{e.g.} Kramers doublets), which would have a degeneracy of $2^{N_{dis}}$.  Another way to differentiate these two scenarios would be to add perturbations, localized near disclinations, breaking all symmetries except $Z_2^I$.\footnote{Unfortunately, it appears impossible to break $Z_2^T$ while preserving $Z_2^I$, without introducing a Monte Carlo sign problem.}  Adding such perturbations will gap out local doublets (which cannot be protected by Ising symmetry alone), but will preserve the vison zero modes.  We note that the presence of local doublets can also be interesting.  For example, following Ref.~\onlinecite{Song2016}, it can be shown that Kramers doublets bound to disclinations are a sign of a non-trivial symmetry protected topological phase, protected by the combination of $D_6$ point group and time reversal symmetry.\cite{pgSPTunpublished}

The vison zero modes should be thought of as a consequence of the symmetry fractionalization of spinons and visons, and in particular of the non-trivial spinon symmetry fractionalization.  We make this connection indirectly:  Any $Z_2$ QSL in the same phase as the one described here can be adiabatically continued so that it is described by the same effective theory and has robust vison zero modes at disclinations, which are a property of the quantum phase.  This $Z_2$ QSL is characterized as a SET phase by the spinon and vison symmetry fractionalization, and only the spinon symmetry fractionalization is non-trivial, so by process of elimination it must be responsible for the vison zero modes.  For example, if we modified the action of Ising symmetry to be trivial on $\sigma^x$ and $\sigma^z$, we would obtain trivial symmetry fractionalization, and nothing would forbid $K_{dis} \neq 0$, so there would be no vison zero modes.  This argument is indirect, and it would certainly be desirable to have a more direct and explicit connection between symmetry fractionalization and vison zero modes, as obtained in Ref.~\onlinecite{Cheng2015} for on-site and translation symmetries.  We have not currently made such a connection, which we leave for future work.

\section{Other properties of the $Z_2$ spin liquid}
\label{sec:other}

Here we use our effective theory to discuss other properties of the $Z_2$ QSL.  Some of these properties are likely challenging to test in quantum Monte Carlo, but may instead be accessible to other numerical approaches.

First, we focus on direct consequences of the non-trivial spinon symmetry fractionalization within the spin liquid phase.  Every state in the single-spinon spectrum is at least doubly degenerate, because a non-degenerate state is not consistent with anticommuting symmetry generators as in Eq.~(\ref{eqn:spinon-nontrivial}).  While the single-spinon spectrum cannot be directly probed, its degeneracies lead to characteristic features in the two-spinon continuum.  Previous works elucidated this structure in cases where translations have non-trivial commutation with other symmetry generators, and found an enhanced periodicity of the two-spinon density of states in crystal momentum.\cite{Wen2002b,WenBook,Essin2014}  Here, acting on a single spinon, translations commute with other generators.  Nonetheless, similar structure is present in the density of states, and can be resolved by point group and Ising quantum numbers.

For simplicity, we focus on $P_y$ and ${\cal I}$ symmetries, and follow  the analysis of Ref.~\onlinecite{Essin2014}.  We consider a two-spinon scattering state $| \psi \rangle$, whose energy is such that single spinon excitations cannot decay (this will always be true near the bottom of the two-spinon continuum).  Without loss of generality, we take $| \psi \rangle$ to be an eigenstate of $P_y$ and ${\cal I}$, with eigenvalues $\sigma_P = \pm 1$ and $\sigma_{\cal I} = \pm 1$, respectively.  The action of symmetry operations on $|\psi \rangle$ factorizes into a product of actions on the two individual spinons, for example
\begin{equation}
P_y | \psi \rangle = P^e_y(1) P^e_y(2) | \psi \rangle \text{.}
\end{equation}
We then consider the effect on $\sigma_P$ of transforming just one of the spinons by the Ising operation,
\begin{equation}
| \psi' \rangle = {\cal I}^e(1) | \psi \rangle \text{.}
\end{equation}
We have
\begin{equation}
P_y  | \psi' \rangle   = P^e_y(1) P^e_y(2) | \psi' \rangle  = - \sigma_P | \psi' \rangle  \text{,}
\end{equation}
and we see that $\sigma_P \to - \sigma_P$.  Now, $| \psi' \rangle $ is an eigenstate with the same energy as $| \psi \rangle$, because ${\cal I}^e(1)$ is a symmetry operation, and the two spinons do not interact in a scattering state.  Similarly, we can find a state of the same energy with $\sigma_{\cal I} \to - \sigma_{\cal I}$.

This discussion can be summarized by defining ${\cal N}_{\sigma_P, \sigma_{{\cal I}} }(\omega)$ to be the density of two-spinon scattering states with $P_y$-eigenvalue $\sigma_P$ and ${\cal I}$-eigenvalue $\sigma_{{\cal I}}$.  We have shown that ${\cal N}_{\sigma_P, \sigma_{{\cal I}} } (\omega)$ is independent of $\sigma_P$ and $\sigma_{\cal I}$.  In particular, the low-energy threshold for the two-spinon continuum is the same in all four symmetry sectors.

Another signature of the spinon symmetry fractionalization involves reduction to
a one-dimensional SPT state.\cite{CYHuang2014, Zaletel2015}  We roll the system
into a cylinder, so that $P_y$ acts effectively as an on-site symmetry of the
one-dimensional system, \emph{i.e.} it does not exchange two ends of the
cylinder.  Then ${\cal I}$ and $P_y$ generate a $Z_2 \times Z_2$ on-site
symmetry, which can protect a single non-trivial SPT phase, the Haldane
phase.\cite{Haldane1983a,Haldane1983b,Gu2009,Pollmann2010,Fidkowski2011,Turner2011,Chen2011,Schuch2011}  In this phase, there are degenerate end states acting on which ${\cal I}$ and $P_y$ anticommute, just as in Eq.~(\ref{eqn:spinon-nontrivial}).
  We consider two different minimally entangled states (MES) of the $Z_2$ spin liquid, that are related by creating a pair of spinons and dragging them to opposite ends of the cylinder.  Equivalently, we can start with one MES and act on it with the string operator transporting a spinon along the cylinder.  One of these MES will be in the trivial $Z_2 \times Z_2$ SPT phase, while the other will be in the Haldane phase, and the difference can be detected via the entanglement spectrum.\cite{Pollmann2010}

Now we turn to the properties of continuous quantum phase transitions that may occur between the  $Z_2$ QSL and nearby conventional ordered phases.  To access such a transition, we can either condense spinons or visons.  The particle that does not condense is gapped at the transition and plays no role there.

To study condensation of visons, we integrate out gapped spinon degrees of freedom, which reduces ${\cal H}_{{\rm eff}}$ to a pure $Z_2$ gauge theory obtained from Eq.~(\ref{eqn:heff}) by dropping the $u$ and $J$ terms, and replacing the gauge constraint with $\prod_{{\sf r}' \sim {\sf r}} \sigma^x_{\sf r r'} = -1$.  Condensation of visons in this theory, which is sometimes referred to as ``odd'' $Z_2$ gauge theory, has been studied before in~[\onlinecite{Moessner2001Ising,Xu2011}].  The simplest possibility, which is driven by nearest-neighbor hopping of visons on the  triangular dual of the honeycomb lattice, is for visons to condense at the Brillouin zone corners ($K$ points), which can lead either to columnar or plaquette valence bond solid (VBS) order, depending on the sign of an anisotropy term.  The transition is in the XY universality class, where the physical VBS order parameter is bilinear in the XY field.    Ref.~\onlinecite{Xu2011} also studied transitions to other types of VBS states that can be driven by adding additional terms to the gauge theory.

In the present context, VBS order is expected for small $J_a$ due to the mapping to the honeycomb lattice quantum dimer model (see Fig.~\ref{fig:phasediagram}).  If this order can be found in quantum Monte Carlo, depending on the type of VBS order, there could be a continuous transition between the VBS and $Z_2$ QSL states.

Turning to condensation of spinons, the first step is to integrate out the gapped vison degrees of freedom.  Before doing that, it is convenient to make a new gauge choice for the action of Ising symmetry, where
\begin{eqnarray}
{\cal I} : \sigma^z_{\sf r r'} &\to& \sigma^z_{\sf r r'} \\
{\cal I} : \tau^z_{\sf r} &\to& - \tau^z_{\sf r} , \qquad {\sf r} \in {\rm A} \\
{\cal I} : \tau^z_{\sf r} &\to& \tau^z_{\sf r} , \qquad {\sf r} \in {\rm B} \text{.}
\end{eqnarray}
The difference from the form given in Table~\ref{tab:1} is that we have ``moved'' (by gauge transformation) the action of ${\cal I}$ from the gauge field to the matter fields.  Integrating out visons corresponds to freezing the magnetic fluctuations of the gauge field, so we set $\sigma^z_{\sf r r'} = 1$, and drop the $K$ and $v$ terms in ${\cal H}_{{\rm eff}}$.  The new gauge choice for Ising symmetry makes this procedure manifestly compatible with the symmetries of the problem.

The effective theory thus becomes two decoupled transverse field Ising models, on the A and B triangular sublattices of the honeycomb lattice.  For simplicity, we assume $J >0$ so that these Ising models are ferromagnetic.  The two Ising models will be coupled by other allowed terms not included in ${\cal H}_{{\rm eff}}$, as  is easily taken into account upon passing to a continuum field theory.  We denote the continuum fields for the two Ising models by $\phi_A$ and $\phi_B$.  To construct a Lagrangian for $\phi_A$ and $\phi_B$, we need to discuss the action of microscopic symmetries.  Both fields change sign under global $Z_2$ gauge transformations.  On the other hand, $\phi_A$ changes sign under Ising symmetry while $\phi_B$ is invariant.  Some of the lattice symmetries (such as $P_y$) exchange A and B sublattices, and therefore take $\phi_A \leftrightarrow \phi_B$.  Taking these symmetries into account, and working in $2+1$-dimensional Euclidean space time with coordinates $\mu = \tau, x, y$, the continuum Lagrangian is
\begin{eqnarray}
{\cal L} &=& \frac{1}{2} \big[ ( \partial_{\mu} \phi_A )^2 + ( \partial_{\mu} \phi_B)^2 \big]  + \frac{m}{2} ( \phi_A^2 + \phi_B^2 ) \\
&+& \lambda ( \phi_A^2 + \phi_B^2 )^2 +  \lambda' \phi_A^2 \phi_B^2  \text{.}
\end{eqnarray}
Here we have included all quadratic terms with two or fewer derivatives, and all quartic terms with no derivatives.

For $\lambda' = - 2 \lambda$, ${\cal L}$ reduces to two decoupled $\phi^4$ field theories, which are constrained by symmetry to have the same parameters.  One can contemplate an Ising $\times$ Ising transition, but the $\phi_A^2 \phi_B^2$ coupling is relevant at this fixed point, so the Ising $\times$ Ising transition can only exist as a multicritical point.

Setting instead $\lambda' = 0$, we have a XY model.  The $\lambda'$ term is a four-fold anisotropy that is known to be irrelevant at the XY critical point (see [\onlinecite{Lou2007}] and references therein).  This suggests that there can be a continuous transition in the XY universality class between the $Z_2$ QSL, where $\langle \phi_A \rangle = \langle \phi_B \rangle = 0$, and an ordered state with a $\phi_A, \phi_B$ condensate.  To establish this conclusively, it would be necessary to consider allowed higher derivative terms and show they are irrelevant.

The nature of the ordered state depends on the sign of $\lambda'$.  For $\lambda' > 0$, the condensate can take on four values, namely
\begin{eqnarray}
\langle \phi_A \rangle = \pm \phi_0 &,& \langle \phi_B \rangle = 0 \\
\langle \phi_A \rangle = 0 &,& \langle \phi_B \rangle = \pm \phi_0 \text{.}
\end{eqnarray}
The overall sign of the condensate is not physical, because it can be changed by a global $Z_2$ gauge transformation, so there are two distinct ground states.  In this phase, Ising symmetry is preserved, but those point group symmetries exchanging the A and B sublattices are broken.  A microscopic realization of this ordering pattern is a density wave of $S^+_{\boldsymbol{r}} S^+_{\boldsymbol{r}'}$ on nearest-neighbor kagome bonds, where 
 $\langle S^+_{\boldsymbol{r}} S^+_{\boldsymbol{r}'} \rangle = c \pm \delta$, with the positive (negative) sign on bonds contained in up-pointing (down-pointing) triangles.
 
For $\lambda' < 0$, up to $Z_2$ gauge transformations there are two distinct states, with
\begin{equation}
\langle \phi_A \rangle = \pm \langle \phi_B \rangle \text{.}
\end{equation}
Here, Ising symmetry is broken, and all lattice symmetries are preserved, so this is the same ferromagnetic state observed by CHM in the XYZh model.\cite{Carrasquilla2015}  There, a first-order transition was found between the QKI regime and the ferromagnetic state.  Our analysis suggests that this transition could potentially be made continuous, and in the XY universality class, by some suitable modification of the XYZh model.

\section{Discussion}
\label{sec:discussion}

In this paper, we derived an effective $Z_2$ gauge theory to explore the hypothesis that the QKI state observed in the XYZh model is a $Z_2$ QSL.\cite{Carrasquilla2015}  In addition to other properties, we found that lattice disclination defects in the $Z_2$ QSL host vison zero modes, which lead to striking observable signatures in the spin susceptibility and entropy.  It would be exciting if these predictions can be tested in future numerical studies of the XYZh model.

The possibility of anyon zero modes at symmetry defects, including flux defects of on-site symmetries and lattice dislocations, has already been pointed out in Ref.~\onlinecite{Cheng2015}.  However, it appears that little attention has been given to such phenomena so far.  In part because anyon zero modes can give rise to striking observable consequences, as we discussed here, further work on this topic may be worthwhile.

\acknowledgments{We are grateful to Roger Melko for a useful discussion.  This research is supported by the U.S. Department of Energy, Office of Science, Basic Energy Sciences (BES) under Award number DE-SC0014415 (Y.-P.~H. and M.H.). Y.-P.~H. is also supported in part by the Taiwan Ministry of Education.}

\appendix

\section{Computation of spinon and vison symmetry fractionalization}
\label{app:frac}

In the main text, symmetry fractionalization was described in terms of operators giving the action of symmetry on a single spinon or vison.  For exactly solvable toric code type models, such operators can be explicitly constructed, and used to compute the symmetry fractionalization of spinons and visons.\cite{Song2015}  Here, using the fact that the QKI $Z_2$ spin liquid has a solvable point that is equivalent to a toric code model, we compute the symmetry fractionalization, largely following Ref.~\onlinecite{Song2015}.

The $Z_2$ gauge theory ${\cal H}_{{\rm eff}}$ of Eq.~(\ref{eqn:heff}) is exactly solvable when $J = v = 0$, because $B_p$ and $\tau^x_{\sf r}$ commute with ${\cal H}_{{\rm eff}}$ and form a complete set of commuting operators.  To make contact with Ref.~\onlinecite{Song2015}, we now exploit the well-known mapping between $Z_2$ gauge theories and toric code models,\cite{kitaev03} which maps the solvable point of the gauge theory to a solvable toric code.

The toric code Hilbert space has a single Ising spin on each link of the honeycomb lattice, for which we write Pauli operators $\mu^z_{\sf r r'}$, $\mu^x_{\sf r r'}$.  The Hilbert space is a tensor product of single-spin Hilbert spaces; there are no gauge constraints.  The mapping between gauge theory and toric code Hilbert spaces is given by
\begin{eqnarray}
\mu^z_{\sf r r'} &=& \tau^z_{\sf r} \sigma^z_{\sf r r'} \tau^z_{\sf r'} \\
\mu^x_{\sf r r'} &=& \sigma^x_{\sf r r'} \text{.}
\end{eqnarray}
It follows that
\begin{equation}
\tau^x_{\sf r} = - \prod_{{\sf r}' \sim {\sf r} } \mu^x_{\sf r r'} \text{,}
\end{equation}
where we used the gauge constraint Eq.~(\ref{eqn:z2gauss}).  From these mappings, it is straightforward to determine the action of symmetry on $\mu^z_{\sf r r'}$ and $\mu^x_{\sf r r'}$.  We have
\begin{eqnarray}
g : \mu^{x,z}_{\sf r r'} &\to& \mu^{x,z}_{g({\sf r}), g({\sf r}') } \\
{\cal T} : \mu^{x,z}_{\sf r r'} &\to& \mu^{x,z}_{\sf r r'} \\
{\cal I} : \mu^{z}_{\sf r r'} &\to& -\mu^{z}_{\sf r r'} \\
{\cal I} : \mu^{x}_{\sf r r'} &\to& \mu^{x}_{\sf r r'} \text{,}
\end{eqnarray}
where $g$ is a $p6m$ space group operation.

The gauge theory Hamiltonian ${\cal H}_{{\rm eff}}$ maps to the toric code Hamiltonian
\begin{equation}
\tilde{{\cal H}}_{{\rm toric}} = u \sum_{\sf r} A_{\sf r} - K \sum_{p} B_p \text{,}
\end{equation}
where $A_{\sf r} \equiv \prod_{{\sf r}' \sim {\sf r} } \mu^x_{\sf r r'}$ and $B_p = \prod_{{\sf r r'} \in p} \mu^z_{\sf r r'}$.  The only difference from the usual toric code on the honeycomb lattice is the sign of the $A_{\sf r}$ term.  We can change this sign by making a basis change, using the  unitary transformation $U = \prod_{\sf r r'} \mu^z_{\sf r r'}$, which sends $\mu^x \to - \mu^x$, and results in
\begin{equation}
{\cal H}_{{\rm toric}} = -u \sum_{\sf r} A_{\sf r} - K \sum_{p} B_p \text{.}
\end{equation}
The action of the symmetry operations on the Pauli operators  remains unchanged in the new basis.

To summarize, we have mapped the problem to the usual toric code model on the honeycomb lattice.  Space group and time reversal act on a trivial way on Pauli operators, but Ising symmetry acts non-trivially on $\mu^z$.  We can therefore anticipate that the subgroup $p6m \times Z^T_2$ has trivial symmetry fractionalization for both spinons and visons, and that any non-trivial part of the symmetry fractionalization must involve the Ising symmetry.  We now outline a more detailed calculation of the symmetry fractionalization, which confirms this expectation.

Before describing the calculation, we first give a more detailed description of what is meant by the spinon (``$e$-particle'') symmetry fractionalization.\cite{Essin2013}  (The description for visons is identical.)  The generators (${\cal I}$, ${\cal T}$, $P_d$, $P_y$, $T_1$ and $T_2$) and relations of the symmetry group are described in Sec.~\ref{sec:symmetry}.  We introduce operators ${\cal I}^e$, ${\cal T}^e$, $P_d^e$, $P_y^e$, $T_1^e$ and $T_2^e$ giving the action of each generator on a single spinon.  These operators obey the same relations as in the symmetry group, but only up to $Z_2$-valued phase factors.  That is,
\begin{eqnarray}
	\left( \mathcal{I}^e \right)^2 &=& \sigma^e_{{\cal I}}  , \qquad
	 \left( \mathcal{T}^e \right)^2 = \sigma^e_{{\cal T}} \\
	 {\cal I}^e {\cal T}^e {\cal I}^e {\cal T}^e &=& \sigma^e_{{\cal I T}} \\
	\left( P_d^e \right)^2&=& \sigma^e_{pd} , \qquad \left( P_y^e \right)^2 = \sigma^e_{py} \\
	\left( P_d^e P_y^e \right)^6 &=& \sigma^e_{pdpy} \\
	T_1^e T_2^e T_1^{e-1}T_2^{e-1} &=&  \sigma^e_{t1t2} \\
	T_1^e P_y^e T_1^{e-1} P^e_y &=& \sigma^e_{t1py} \\
	T_1^e {\cal I}^e T_1^{e-1} {\cal I}^{e-1} &=& \sigma^e_{t1 {\cal I}} \\	
	P_d^e {\cal I}^e P_d^{e-1} {\cal I}^{e-1} &=& \sigma^e_{pd {\cal I}} \\	
	P_y^e {\cal I}^e P_y^{e-1} {\cal I}^{e-1} &=& \sigma^e_{py {\cal I}} \\	
	T_1^e {\cal T}^e T_1^{e-1} {\cal T}^{e-1} &=& \sigma^e_{t1 {\cal T}} \\	
	P_d^e {\cal T}^e P_d^{e-1} {\cal T}^{e-1} &=& \sigma^e_{pd {\cal T}} \\	
	P_y^e {\cal T}^e P_y^{e-1} {\cal T}^{e-1} &=& \sigma^e_{py {\cal T}} \\	
	T_2 &=& P_dT_1P_d \\
	T_2P_y &=& P_yT_1T_2^{-1}  \text{.}
\end{eqnarray}
Here, each $\sigma^e$ parameter can be either $+1$ or $-1$.  The generators can be redefined by a minus sign without affecting any physical properties; for example, $T_1^e \to - T^1_e$ is an allowed redefinition.  The $\sigma^e$ parameters are invariant under such redefinitions, and specifying all 14 $\sigma^e$'s gives the spinon fractionalization class, which is an element of $H^2(G, Z_2) \simeq (Z_2)^{14}$.  The last two relations have no $\sigma^e$ parameters because the generators can be suitably redefined to remove any phase factors.

We find that all the $\sigma^e$'s are unity, except $\sigma^e_{py {\cal I}} = -1$.  Since $P_y$ is the only $p6m$ generator that exchanges the A and B honeycomb sublattices, it follows that ${\cal I}^e$ anticommutes with precisely those $p6m$ operations exchanging the two sublattices, while it commutes with operations taking A to A and B to B.  For the vison, we find that all the $\sigma$ parameters are unity; that is, the vison fractionalization class is trivial.  

We now describe how the spinon symmetry fractionalization is computed.  We omit the computation of the vison symmetry fractionalization, as 
it can be straightforwardly obtained by the same means.  We first follow Ref.~\onlinecite{Song2015} to obtain the $\sigma^e$'s involving only unitary operations.  We then determine the $\sigma^e$'s involving time-reversal by a different argument.

In the ground state of the toric code, $A_{\sf r} = 1$, and spinons reside at honeycomb sites with  $A_{\sf r} = -1$.  To create a pair of spinons at ${\sf r}_1$ and ${\sf r}_2$, we act on the ground state with a string operator ${\cal L}^e_s$, where $s$ is a path of links on the lattice joining ${\sf r}_1$ and ${\sf r}_2$, and ${\cal L}^e_s$ is a product of $\mu^z_{\sf r r'}$ over this path.  Such string operators also transport a spinon from one site to another.  We consider two-spinon states
\begin{equation}
| \psi_e (s) \rangle = {\cal L}^e_s | \psi_0 \rangle \text{,}
\end{equation}
where for simplicity we assume the ground state $|\psi_0 \rangle$ is invariant under all symmetry operations.  The state
$| \psi_e (s) \rangle$ only depends on the endpoints of the path $s$.

We let ${\mathfrak g}$ be a unitary element of the symmetry group, realized by the operator $U_{\mathfrak g}$.  Ref.~\onlinecite{Song2015} showed that we can find operators $U^e_{\mathfrak g}({\sf r})$ giving the action of ${\mathfrak g}$ on the spinon at ${\sf r}$, satisfying
\begin{equation}
U_{\mathfrak g}  | \psi_e (s) \rangle  = U^e_{\mathfrak g}({\sf r}_1) U^e_{\mathfrak g}({\sf r}_2) | \psi_e (s) \rangle  \text{.}
\end{equation}
In general,
\begin{equation}
U^e_{\mathfrak g}({\sf r}) = f_{\mathfrak g}({\sf r}) {\cal L}^e_{s_{\mathfrak g} ({\sf r}) } \text{,}
\end{equation}
where $f_{\mathfrak g}( {\sf r} ) \in \{ \pm 1 \}$, and $s_{\mathfrak g}({\sf r})$ is a path joining ${\sf r}$ to ${\mathfrak g} ( {\sf r })$.  Only the action of $U^e_{\mathfrak g}({\sf r})$ on states of the form $| \psi_e(s) \rangle$ is of any consequence, and different choices of  $U^e_{\mathfrak g}({\sf r})$ having the same action are considered equivalent.
For the toric code model we are considering here, the choice of path $s_{\mathfrak g}({\sf r})$ (for fixed endpoints) does not affect the action of $U^e_{\mathfrak g}({\sf r})$ on $ | \psi_e(s) \rangle$, so we can completely specify $U^e_{\mathfrak g}({\sf r})$ by $f_{\mathfrak g}({\sf r})$.

It was shown in Ref.~\onlinecite{Song2015} that the operators $U^e_{\mathfrak g}({\sf r})$ have a unique action on states $|\psi_e(s) \rangle$
 up to projective transformations
$U^e_{\mathfrak g}({\sf r})  \to \lambda( {\mathfrak g} ) U^e_{\mathfrak g}({\sf r})$, where $\lambda( {\mathfrak g}) \in \{ \pm 1\}$.  Working in terms of generators and relations, these transformations simply express the freedom to redefine $U^e_{\mathfrak g}({\sf r})$ by a minus sign for ${\mathfrak g}$ a generator.  

The $\sigma^e$ parameters can then be calculated by acting with appropriate products of $U^e_{\mathfrak g}({\sf r})$ on a state $| \psi_e (s) \rangle$.  For example, to calculate $\sigma^e_{py}$, we write
\begin{equation}
U^e_{P_y}[ P_y( {\sf r}_1) ]  U^e_{P_y} ( {\sf r}_1) | \psi_e (s) \rangle  = \sigma^e_{py}  | \psi_e (s) \rangle \text{,}
\end{equation}
and evaluate the left-hand side.

As mentioned above, the operators $U^e_{\mathfrak g}({\sf r})$ are completely specified by $f_{\mathfrak g}( {\sf r})$.  It is not difficult to see that
\begin{equation}
f_g ( {\sf r} ) = 1 \text{,}
\end{equation}
for $g$ any space group operation, including the generators $T_1$, $T_2$, $P_y$ and $P_d$.  For the Ising symmetry, we have
\begin{equation}
f_{{\cal I}} ( {\sf r} ) = \left\{ \begin{array}{ll}
+1 , & {\sf r} \in {\rm A} \\
-1 , & {\sf r} \in {\rm B} 
\end{array}\right.  \text{.}
\end{equation}
With this information, it is straightforward to follow the prescription described above and determine all the $\sigma^e$ parameters not involving time reversal.  We find that they are all equal to unity except $\sigma^e_{py {\cal I}} = -1$.

It would not be difficult to extend the formalism of Ref.~\onlinecite{Song2015} to incorporate time reversal, but to maximize efficiency and minimize the introduction of new formalism, we use a different set of arguments to determine the remaining five $\sigma^e$ parameters involving ${\cal T}$.  All of these parameters are associated with particular symmetry-protected degeneracies in the single-spinon spectrum that we now show are not present.

First, we consider $\sigma^e_{\cal T}$.  If this parameter were $-1$, spinons would be Kramers doublets, but a spinon localized on the lattice site ${\sf r}$ is clearly non-degenerate, implying $\sigma^e_{\cal T} = 1$.  Similarly, we must also have $\sigma^e_{{\cal I T } } = 1$, since otherwise ${\cal I}^e$ and ${\cal T}^e$ would anticommute, and a spinon localized to ${\sf r}$ would have at least a two-fold degeneracy.

The remaining $\sigma^e$ parameters involving time reversal are $\sigma^e_{t1 {\cal T}}$,  $\sigma^e_{pd {\cal T}}$ and $\sigma^e_{py {\cal T}}$.  These parameters involve the space group generators, which in general are not symmetries of a spinon localized at ${\sf r}$.  Instead, it is convenient to consider single-spinon plane wave states.  We perturb the toric code Hamiltonian by adding $\delta H = -\tilde{h} \sum_{\sf r r'} \mu^z_{\sf r r'}$, which breaks $Z_2^I$ but preserves $p6m \times Z_2^T$.  This term is a nearest-neighbor hopping for spinons, and results in a spinon dispersion with a non-degenerate minimum at $\boldsymbol{k} = 0$.  If any of $\sigma^e_{t1 {\cal T}}$,  $\sigma^e_{pd {\cal T}}$ or $\sigma^e_{py {\cal T}}$ were equal to $-1$, time reversal would anticommute with some space group operations, which is inconsistent with having a non-degenerate single-spinon energy eigenstate.  Therefore, all of the $\sigma^e$ parameters involving time reversal are equal to $+1$.

\bibliographystyle{apsrev4-1}
\bibliography{QKI.bib}

\end{document}